\documentclass[twocolumn,twoside]{article}
\usepackage{graphics}
\newcommand{\hb}{\\ \hspace*{2ex}}

\begin{document} %%%%%%%%%%%%%%%

\title{Some evolutionary aspects of the binary stellar systems containing
neutron star.}

\author{O.M.\,Ulyanov$^{1}$, S.M.\,Andrievsky$^{2}$, V.F.\,Gopka$^{2}$, A.V. Shavrina$^{3}$\\[2mm] %English only
\begin{tabular}{l}
 $^1$  Institute of Radio Astronomy of the National Academy of Sciences
of Ukraine\hb Chervonoprapona str. 4, Kharkov 61002, Ukraine, {\em
oulyanov@rian.kharkov.ua}\\[2mm]
 $^2$ Department of Astronomy,
      Odessa National University\hb
 T.G. Shevchenko Park, Odessa 65014 Ukraine,
 {\em scan@deneb1.odessa.ua; gopka.vera@mail.rus}\\[2mm]
$^3$ Main Astronomical Observatory of NASU, Zabolotnogo str.
27,\hb Kyiv 03680, Ukraine, {\em shavrina@mao.kiev.ua}\\[2mm]
\end{tabular}
% no tabular, if one affiliation only
}
\date{}
\maketitle

\section*{Abstract} %%%%%%%%%%%%%%%

The obvious lack of the  binary stellar systems that contain
neutron stars (NS) is observed at present. Partly it is caused by
the fact that it is very difficult to detect neutron star in
a binary system if this relativistic component does not manifest
itself as a radio pulsar. Among 1879 pulsars that are listed
in the ATNF pulsar catalogue, only 141 pulsars are known to be the
companions in binary systems. Only 81 objects having median mass
estimation of more than 0.2 $M_{\odot}$ constitute the binary
systems with pulsars.  Nevertheless, such systems should be much more
numerous and their investigation is of the great interest because
thier structure and evolution can certainly help in our understanding
of many unique properties that are seen in some stars.

\section{Introduction} %%%%%%%%%%%%%%%

Some difficulties of a detection of the NS as sources of the radio
emission [1] are discussed in Refs. [2, 3, 4, 5, 6]. Analysis of
the different stages of the close binary system evolution with a NS,
as one of the companion stars, shows that at the early stage of this
evolution the ultra-relativistic electron-positron plasma produced
and ejected by NS can hit the atmosphere gas of its companion normal
star. At the same time, an usual stellar wind of the normal star will
eventually screen NS magnetic field. During some period both winds
from NS and normal star will create sources of X-ray emission located
at the surfaces of both companions.

The detection of the annihilation line at 511 keV from such systems
could be considered as observed manifestation of the above mentioned
interaction between NS and normal star in the binary system, in addition
to the detection of X-ray radiation. An appearance of the chemical
composition anomalies in the companion star atmosphere can also be possible
in this case.

There are papers [7, 8, 9, 10, 11] in which two most probable
evolutionary scenarios are proposed. The first scenario is
connected with an expansion of the donor under the action of the
X-ray and gamma-ray radiation generated by the NS (acceptor) [11].
In this scenario, the generation of X-ray and gamma-ray radiation
by the NS is a result of accretion of the wind of a companion star.
The expansion of the upper layers of the donor in this model is caused
by the energy absorption of the NS X-ray and gamma-radiation in
the relatively deep layers of the donor. This model implies that
donor fills its entire Roche lobe and its material is lost through the
first Lagrangian point.

According to this scenario, under the influence of the hard
radiation emitted by NS, the donor (companion normal star) passes
through the cyclic stages of the orbital period change. At the
same time, as a whole, the orbital period decreases, similarly to
that as this occurs in cataclysmic binary stars [9, 10, 11].

Moreover, this scenario predicts the secular increase of the
semi-major axis due to an increase of the Roche lobe radius of the
donor, and corresponding increase of the mass exchange rate. The
most contrasting manifestation of this effect should be observed
in the range of the hard radiation fluxes from $10^{10}$ to
$10^{12}$~$ergs/(cm^{2} \cdot sec)$, and for the range of the
donor masses from 0.1 to $2.0 M_{\odot}$. This process must occurs
in the so-called Jeans mode [12], when for a long time the
conditions $A_{m} \cdot (M_{1} + M_{NS}) = const$ are satisfied,
where $A_{m}$ is the size of the semi-major axis, $M_{1}$ is the
mass of the donor, $M_{NS}$ is the mass of the NS (acceptor).

The second scenario describes evolutionary evaporation of the
donor due to the presence of the so-called induced stellar wind.
In this case, it is assumed that the hard radiation of acceptor
heats the relatively thin upper layers of the donor, which have
the low density. Cooling in this scenario is achieved due to an
induced stellar wind that leads to the gradual evaporation of the
donor. In contrast to the first scenario, in this case the size of
the semi-major axis of the binary system will decrease with a time.
At the early stages of evolution, this decrease is caused by the
loss of the binary system moment of inertia due to the stellar
wind. However, in the case of an extremely close binary system
the gravitational radiation losses must play a more significant role.
The qualitative analysis of the interaction in the close binary system
with a young NS, as one of the components [12, 13, 14, 15],
shows that at the early stage of this evolution heating of the
upper atmosphere of the companion can be achieved not only due to
the NS hard radiation, but also due to its ultra-relativistic
electron-positron plasma ejection. In this case, plasma ejected
by NS can hit the atmosphere layers of the companion star. The depth
of the heating is supposed to be not uniform. Such an ununiformity
can be connected with a concrete geometry of the magnetic field in
the donor atmosphere [16]. For example,as far as the synchrotron
energy losses are proportional to the component of the magnetic field
perpendicular to the charge flow, zones near the magnetic poles will
be heated at larger depths compared to other zones of the donor
atmosphere. Let us note that ionization will be the basic source
of energy loss for the ultra-relativistic electrons and positron,
which have relatively low kinetic energies [17]. These losses
will occur in the upper and relatively thin layers of a companion
star [16].

Detailed analysis of different types of the losses of the
ultra-relativistic electron-positron plasma in the companion
atmosphere was performed in [16]. The losses caused by the
inverse Compton effect, that takes place near the companion
(because of an interaction of the electron-positron ultra-relativistic
plasma with optical photons), are not essential in comparison
with other basic types of the energy losses [16, 17]. During this
period of binary system evolution both winds create the sources
of X-ray radiation on the surfaces of companion stars. As it was
already mentioned, observational manifestation of such an interaction
(besides the X-ray radiation) might be detection of the annihilation
line at 511 keV [16, 18]. An appearance of the chemical anomalies in
the atmosphere of the companion is also possible (for instance, some
lines of the unstable radioactive isotopes were registered in spectra
of some stars) [19, 20].

An interaction of two objects in the system containing NS can lead
to the gradual screening of the NS magnetic field, and corresponding
weakening of the ultra-relativistic plasma flow. This may cause the
"switch off" of the X-ray and gamma-ray sources of radiation in the
atmosphere of the normal companion star. At the same time, X-ray emission
from the NS surface can become stronger. Further evolution strongly depends
on the diffusion of the NS magnetic field through the acquired accretion
crust.

\section{Observation Data} %%%%%%%%%%%%%%%

Let us examine the observational facts available at present that characterize
evolution of the binary systems which contain NS.

There is ATNF catalogue [1], where important properties are gathered for
all pulsars. In this catalogue, the listed pulsars are the members of binary
systems with primary companions of different types. In fact these are close
binary systems. After examining of the whole amount of available data in ATNF
catalogue, one can make the following conclusions:\\

1) in the binary systems the secular decrease of the magnetic field strength
on the pulsar surface is observed (Fig. 1.);\\

2) the well known effect of the evolutionary decrease of the pulsar rotation
period is observed;\\  %(Fig. 2.)

3) the general stabilization of the pulsar kinematic parameters connected with the
decreasing of the magnitude of $\dot{P}$ and with an increase of the so-called characteristic
age $Age = P/(2 \dot{P})$, (where $Age$ is characteristic age of pulsar, $P$ is the period of
the rotation of NS around its own axis, $\dot{P}$ is the first-order derivative of the P)
is observed (Fig. 2.);\\

4) on average, for the lower masses of the normal companions, the decrease of the
semi-major axis is observed (excluding the companion mass range from $0.1 M_{\odot}$
to $2M_{\odot}$) (Fig. 3.);\\

5) the mass decrease of the normal companion as a function of an increase of the
pulsar characteristic age is observed (Fig. 4.).\\

\begin{figure}
%\resizebox{4.26cm}{!}  % Fig_01_Bs_Age.eps
\resizebox{\hsize}{!} {\includegraphics{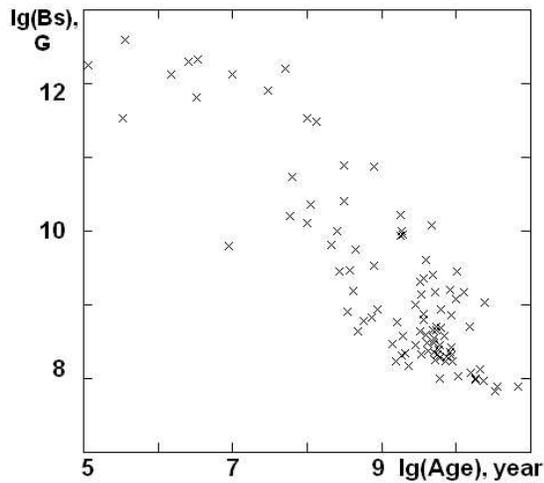}}
\label{hh} \caption{Distribution of the magnitude of the surface
strength of magnetic field ($B_{s}$)
      of pulsars in binary systems as a function of their characteristic ages ($Age = P/(2 \dot{P})$).}
\end{figure}

\begin{figure}
%\resizebox{4.26cm}{!}  % Fig_02_Pdot_Age.eps
\resizebox{\hsize}{!} {\includegraphics{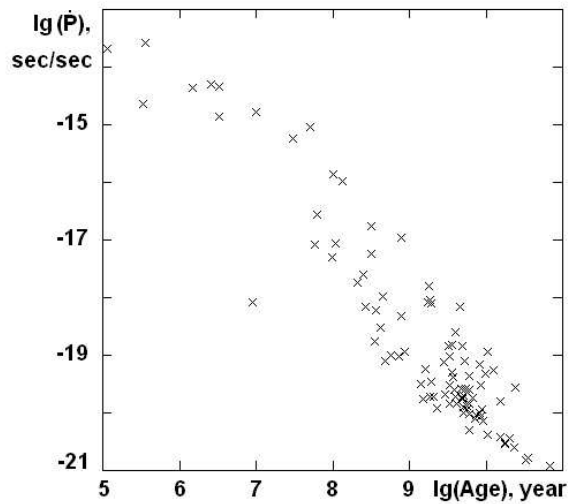}}
\label{hh} \caption{Distribution of the first derivatives for the
pulsar rotation periods ($\dot{P}$)
       as a function of the pulsar characteristic age.}
\end{figure}

\begin{figure}
%\resizebox{4.26cm}{!}  % Fig_03_Amaj_Med_mass.eps
\resizebox{\hsize}{!} {\includegraphics{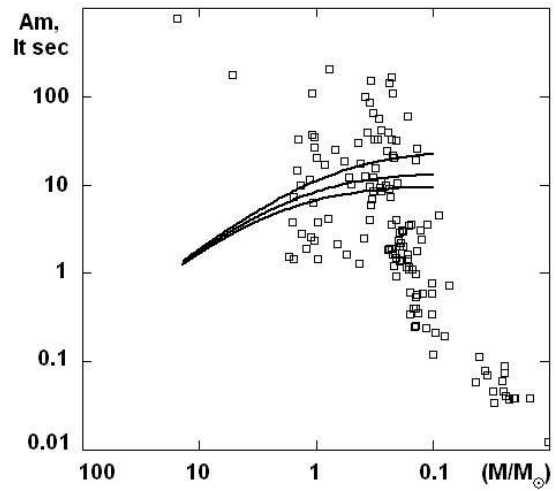}}
\label{hh} \caption{Distribution of the magnitude semi-major axis
       of the binary systems ($A_{m}$) depending on the median mass
        of the normal companions. Solid lines illustrate qualitative behaviour
        under the Jeans mode conditions corresponding to $0.8M_{\odot}, 1.4M_{\odot},
        2.0M_{\odot}$ pulsar masses (respectively from the top to bottom).}
\end{figure}

\begin{figure}
%\resizebox{4.26cm}{!}  % Fig_04_Med_mass_Age.eps
\resizebox{\hsize}{!} {\includegraphics{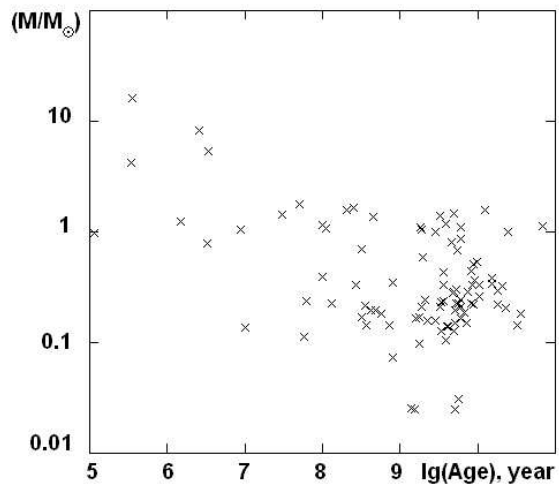}}
\label{hh} \caption{Distribution of the median mass of the normal
companions ($M/M_{\odot}$)
       as a function of the pulsar characteristics.}
\end{figure}

\section{Possible Interpretation of the Observation Data} %%%%%%%%%%%%%%% 3

It is easy to interpret the first three dependencies mentioned
above (items 1-3 and Fig. 1-2. The secular decrease of the
magnetic field strength on the pulsar surface (Fig. 1.) most
probably is connected to its screening with the material falling
on the pulsar surface from its normal companion. This possibility
was indicated in a number of papers (for example, see [21] and
references therein). Here we will only note that the rate of
accretion must correlate with the intensity of the hard radio
emission, produced with the same particles via the warming-up of
pulsar surface up to extremely high temperatures. The second
aspect is the fact that the magnetic field will be preserved
inside the core of pulsar/NS for a long time, but when the
accretion stops the magnetic field must float to its surface as a
result of diffuse processes. The threshold value of magnetic
induction on the NS surface, above which we observe the NS as a
radio pulsar, according to the ATNF catalog data corresponds to
$B_{s} > 6.7 \cdot 10^{7}$ Gauss.

Similar explanation involves the evolutionary decrease of the
pulsar rotation periods. In this case (Fig. 2.), the so-called
effect of the slight twist occurs, when the donor twists of the
pulsar, increasing its torque. The effect is connected with an
increase of the conservatism of such kind of the binary system
(Fig. 2.). This is confirmed by the decrease in the first-order
derivative of the pulsar rotation period. When the magnetic field
strength on the pulsar surface decreases, the corresponding
decrease in the kinematic losses of pulsar (caused either by
magneto-dipole emission or by current losses) takes place. It is
quite probable that the accretion on pulsar surface is stabilized
or completely discontinued as pulsar characteristic age increases.

Items 4, 5 (Fig. 3-4) can be grouped together due to the fact that
total mass of both companions, sizes of semi-major axes and
orbital periods are coupled via the third Kepler's law. Although
the general trend (Fig. 4) probably indicates to a decrease of the
mass of the NS companion in the binary systems, the behavior of
these systems in the range of the masses $M_{1}\in [0.1M_{\odot};
2M_{\odot}$] should be considered in more detail. For example, in
the range of masses $M_{1} \in [0.1 M_{\odot}; 2 M_{\odot}]$ the
companion normal star can loose very rapidly part of its mass as a
result of the tidal destruction, directed explosions, or a series
of the directed micro explosions/bursts.

\section{Conclusions} %%%%%%%%%%%%%%%

The analysis of the presented data shows that in the close binary
systems containing NS, the secular decrease of the magnetic field
strength on the NS surface is observed. This decrease of the
magnetic field strength most probably is connected with the
accretion from the companion star on the NS surface.
A threshold below which the NS/pulsar cannot be observed as a
radio source is about ~ $6.7 \cdot 10^{7}$ Gauss.

The observed secular decrease of the first derivative of the
rotation period of the pulsars most likely corresponds to the fact
that between the pulsar and its companion star a deep negative
feedback is established, which effectively  suppresses any
fluctuations in the rate of the accretion on the NS surface and/or
fluctuation in the luminosity of the hard radiation.

The evolutionary decrease of the semi-major axis of the binary
systems and the simultaneous decrease of the mass of the normal
star companions over the large time intervals testifies in a favor
of the hypothesis about evolutionary evaporation (or destruction
with the subsequent evaporation) of the companion star.

This hypothesis is also supported by the presence of numerous
single millisecond pulsars with a large characteristic age, whose
companion stars are not seen.

In the range of companion masses from $ 0.1 M_{\odot}$ to $2
M_{\odot}$, the most complex manifestation of the evolutionary
phenomena in the binary systems are observed. It is quite probable
that in this mass range the Jeans accretion mode operates (see
Fig. 3).

%%%%%%%%%%%%%%%%%%%%%%%%%%%%%%%%%%
%% thebibliography environment %%
%%%%%%%%%%%%%%%%%%%%%%%%%%%%%%%%%

%%%%%%%%%%

\end{document}